\documentclass[pra,preprint,aps,floats]{revtex4}
\usepackage{graphicx}
\usepackage{amsmath}
\newcommand{\xr}{x_{ref}}
\newcommand{\ex}{e^{-\sqrt{a\epsilon}\xr}}
\newcommand{\exe}{e^{-\sqrt{\epsilon}\xr}}
\begin{document}
\title{Improving the Convergence of an  Iterative Algorithm Proposed By Waxman}
\author{W. A. Berger\footnote{E-mail wb@wberger.com} and H. G. Miller\footnote{E-Mail: hmiller@maple.up.ac.za}}
\affiliation{Department of Physics, University of Pretoria, Pretoria 0002, South Africa}

\begin{abstract}
In the iterative algorithm recently proposed by Waxman for solving  eigenvalue problems, we 
point out that the convergence rate may be improved. For many non-singular symmetric potentials which vanish asymptotically, 
a simple analytical relationship between the coupling constant of the potential and the ground state eigenvalue is obtained which can be used to make the algorithm more efficient. 

\noindent
PACS\ 03.65.Ge,\ 02.60.Lj 
\end{abstract}
\maketitle
Recently, Waxman\cite{W98} has proposed a convergent  iterative algorithm for obtaining solutions of the eigenvalue problem, which does not involve a matrix diagonalization. For operators which possess a continuum as well as a set of bound states this is a most advantageous\cite{AMP06}.  In the case of  ground state, for example, the eigenenergy, $\epsilon$, is determined numerically  as a function of the coupling constant of the potential, $\lambda$, and inverted to yield the $\epsilon$ corresponding to the required value of $\lambda$. The convergence rate of the algorithm, therefore, depends on two factors: the number of iterations required to find an eigensolution for a particular choice of $\epsilon$; and the number of times this must be repeated in order to determine the value of $\epsilon$ which corresponds to the desired value of $\lambda$. In this regard we wish to point out that for certain potentials a simple relationhip between $\lambda$ and $\epsilon$ exists which may used to improve the convergence rate.       

In the Waxman algorithm , eigenpairs are 
determined as functions of the strength of the potential in the following 
manner.  For simplicity consider a one-dimensional eigenvalue equation\cite{W98}
\begin{equation}
[-\partial^2_{x} -\lambda V(x)] u(x)=-\epsilon u(x) \label{heq}
\end{equation}
\begin{equation}
\lim_{|x|->\infty}u(x)=0
\end{equation}
where $\partial_{x}=\frac{\partial}{\partial_x}$;   $\lambda > 0$ is the 
strength parameter of the attractive potential ($\lambda$V(x) $>$ 0 and 
V(x)$\rightarrow 0 \ as \ |x| \rightarrow  \infty$) and the energy eigenvalue, 
$-\epsilon $ (with $\epsilon  >  0$), is negative and corresponds to a bound 
state. Using Green's method a solution to eq(\ref{heq}) is given by
\begin{equation}
u(x)= \lambda\int^\infty_{-\infty}G_\epsilon(x-x') V(x')u(x')dx' \label{G} 
\end{equation}
where the  Green's function $G_\epsilon(x)$ satisfies
\begin{equation}
[-\partial^2_{x} + \epsilon]G_\epsilon(x)=\delta(x)
\end{equation}
\begin{equation}
\lim_{|x|->\infty}G_\epsilon(x)=0.
\end{equation}
Normalizing u(x) at an arbitrary $x_{ref}$ 
\begin{equation}
u(x_{ref})=1
\end{equation}
allows $\lambda$ to be written as (see eq(\ref{G}))
\begin{equation}
\lambda=\frac{1}{\int G_{\epsilon}(x_{ref}-x')V(x')u(x') dx'} \label{l}
\end{equation}
which can then be used to eliminate $\lambda$ from eq(\ref{G})
\begin{equation}
u(x)=\frac{\int^\infty_{-\infty}G_\epsilon(x-x') V(x')u(x')dx'}{\int 
G_{\epsilon}(x_{ref}-x')V(x')u(x') dx'}\label{u}.
\end{equation}
Using equations (\ref{l}) and (\ref{u}), $\lambda$ can be determined as a function 
of $\epsilon$ in the following manner. For a particular choice of $\epsilon$ 
eq(\ref{u}) can be 
iterated
\begin{equation}
u_{n+1}(x)=\frac{\int^\infty_{-\infty}G_\epsilon(x-x') 
V(x')u_n(x')dx'}{\int G_{\epsilon}(x_{ref}-x')V(x')u_n(x') dx'} 
\label{iu}
\end{equation}
until it converges and $\lambda$ can then be determined 
from eq(\ref{l}). Repeating for different values of $\epsilon$ yields a set of 
different values of the potential strength $\lambda$. When enough points have 
been determined, a simple interpolation procedure can be used to determine the dependence of $\epsilon$ on $\lambda$.

On the other, a simple relationship between $\lambda$ and $\epsilon$ can be obtained
for non-singular symmetric potentials which vanish asymptotically 
in the following manner.  Note that such the eigensolutions of such potentials have good parity. Taking the limit of eq(\ref{heq}) as $x\rightarrow\xr$ yields
\begin{equation}
\lambda=\frac{\lim_{x->\xr}(-\partial_x^2u(x))+\epsilon}{V(\xr)} \label{laml}
\end{equation}
since $u(\xr)=1$.
If the potential is deep enough the ground state eigenvalue, $\epsilon$, is not small.  In this case it is reasonable to let  
 \begin{equation}
 u(x)=f(x)e^{-\sqrt{a\epsilon}x} \qquad x\geq 0 \label{ua}
 \end{equation}
where  a is constant and f(x)  satisfies
\begin{equation}
u(x) \underset{x->\infty}{\rightarrow} 0.
\end{equation}
  The only dependence of u(x) on $\epsilon$ is in the exponential tail of the wave function and is neglected in f(x) which generally has a polynomial structure. For small values of $\epsilon$   neglecting this dependence is in most cases not possible.   
Then
\begin{eqnarray}
\lim_{x->\xr}( -\partial_x^2u(x))&\rightarrow&f''(\xr)\ex- 2\sqrt{a\epsilon}f'(\xr)\ex+a\epsilon\\
&\rightarrow& a_1\exe+ a_2\sqrt{\epsilon}\exe+a_3\epsilon \label{lim}
\end{eqnarray}
where $f'=\frac{d f}{dx}$.
 Combining eqs(\ref{laml}) and (\ref{lim}) yields 
\begin{equation}
\lambda=a_1\exe+ a_2\sqrt{\epsilon}\exe+a'_3\epsilon \label{lame} 
\end{equation}

\begin{figure}
\centering
\begin{center}
\includegraphics[scale=0.5,angle=270]{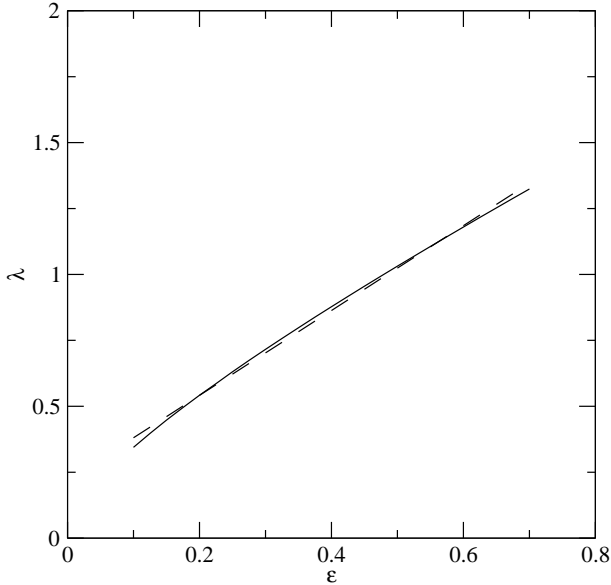}
\end{center}
\label{le}\caption{The coupling constant $\lambda$  as a function of the ground state energy  $\epsilon$ for the lowest-lying even parity eigensolution of the inverted Gaussian potential (solid curve). The coupling constant for the potential is 1.  The dashed curve is a fit to the data given by
$\lambda=.21972+1.6087\epsilon$  }
\end{figure}

Now for $\xr=0$ and potentials which are symmetric about 0, $f'(0)=0$ ($\Rightarrow a_2=0$) for the even parity solutions. In this case one obtains a simple linear dependence of $\lambda$ on $\epsilon$ involving two unknown coefficients. In the more general case or for  a different choice of $\xr$,
 three values of $\lambda$  must be calculated numerically  for three different choices of $\epsilon$ to determine the three coefficients in eq(\ref{lame}) and   the dependence of
 $\lambda$ on $\epsilon$.  Care must be taken that eigensolutions used to determine $\lambda$ are well converged.  Numerical errors  may lead an incorrect dependence and require additional numerical determinations of $\lambda$.
 
 In order to demonstrate the linear dependence of $\lambda$ on $\epsilon$, we have performed a calculation  of lowest-lying even parity eigensolution of the inverse Gaussian potential\cite{AMP06}
\[
V(x)= e^\frac{-x^2}{2}.
\]
Note that $\xr=0$ and $\lambda$=1. For each choice of $\epsilon$, $\lambda$ has been determined from eq(\ref{l}) after the eigensolution, u(x), has converged. 
In figure 1 the linear dependence is clearly 
demonstrated and confirms numerically the existence of a simple linear relationship between $\lambda$ and $\epsilon$ as long as $\epsilon$ is not small.

\vspace{10mm}
\noindent
Acknowledgment\\
Discussions with A. R. Plastino and R. A. Andrew are gratefully acknowledged. 

\end{document}